\documentclass[conference]{IEEEtran}
\IEEEoverridecommandlockouts
\usepackage{cite}
\usepackage{amsmath,amssymb,amsfonts}
\usepackage{graphicx}
\usepackage{textcomp}
\usepackage{xcolor}
\usepackage{multirow}
\usepackage{algorithm}
\usepackage{algpseudocode}
\usepackage{subfig}

\def\BibTeX{{\rm B\kern-.05em{\sc i\kern-.025em b}\kern-.08em
    T\kern-.1667em\lower.7ex\hbox{E}\kern-.125emX}}
\begin{document}

\title{\textbf{TNNGen}: Automated Design of Neuromorphic Sensory Processing Units for Time-Series Clustering}

\author{
\IEEEauthorblockN{
    Prabhu Vellaisamy\IEEEauthorrefmark{1},
    Harideep Nair\IEEEauthorrefmark{1},
    Vamsikrishna Ratnakaram,
    Dhruv Gupta, and
    John Paul Shen
  }
  \IEEEauthorblockA{Electrical and Computer Engineering Department, Carnegie Mellon University\\
  \{\textit{pvellais, hpnair, vratnaka, dhruvgup, jpshen}\}\textit{@andrew.cmu.edu}
  }
  \IEEEcompsocitemizethanks{\IEEEcompsocthanksitem\IEEEauthorrefmark{1}Both authors contributed equally to this work.}
}

\maketitle

\begin{abstract}
Temporal Neural Networks (TNNs), a special class of spiking neural networks, draw inspiration from the neocortex in utilizing spike-timings for information processing. Recent works proposed a microarchitecture framework and custom macro suite for designing highly energy-efficient application-specific TNNs. These recent works rely on manual hardware design, a labor-intensive and time-consuming process. Further, there is no open-source functional simulation framework for TNNs. This paper introduces \textit{TNNGen}, a pioneering effort towards the automated design of TNNs from PyTorch software models to post-layout netlists. TNNGen comprises a novel PyTorch functional simulator (for TNN modeling and application exploration) coupled with a Python-based hardware generator (for PyTorch-to-RTL and RTL-to-Layout conversions). Seven representative TNN designs for time-series signal clustering across diverse sensory modalities are simulated and their post-layout hardware complexity and design runtimes are assessed to demonstrate the effectiveness of TNNGen. We also highlight TNNGen's ability to accurately forecast silicon metrics without running hardware process flow.

\end{abstract}


\section{Introduction and Background}

Deep neural networks (DNNs) \cite{lecun2015deep} have emerged as the de facto technology for artificial intelligence (AI), even surpassing human-like sensory processing capabilities. However, this impressive progress comes with an exponential surge in compute demands and energy consumption, raising concerns about long term sustainability of this trend \cite{openai, thompson2020computational, numenta}. Temporal Neural Networks (TNNs), a special class of spiking neural networks (SNNs) employing spike time-based processing with close adherence to biological plausibility \cite{smith2018space, smith2017space, smith2020temporal}, offer a promising alternative path for AI compute, with potential for orders of magnitude improvements on energy efficiency. 

Recent research \cite{chaudhari2021unsupervised} demonstrates that single-layered TNNs excel in unsupervised time-series clustering \cite{aghabozorgi2015time, zhang2018salient, ma2019learning} and are amenable for resource-constrained edge devices. Further works in advancing TNN research include a microarchitecture framework for TNN implementation in 45nm CMOS \cite{nair2021microarchitecture}, and augmenting the ASAP7 process design kit (PDK) \cite{clark2016asap7} with TNN-tailored custom macros for improved energy-efficiency \cite{nair2022tnn7}. Recently, the idea of creating an end-to-end framework that can automate the design of specialized TNN chiplets for online sensory processing applications was suggested in \cite{vellaisamy2022towards}.



\begin{figure}[t]
\centering
\includegraphics[width=0.99\columnwidth, height=14cm]{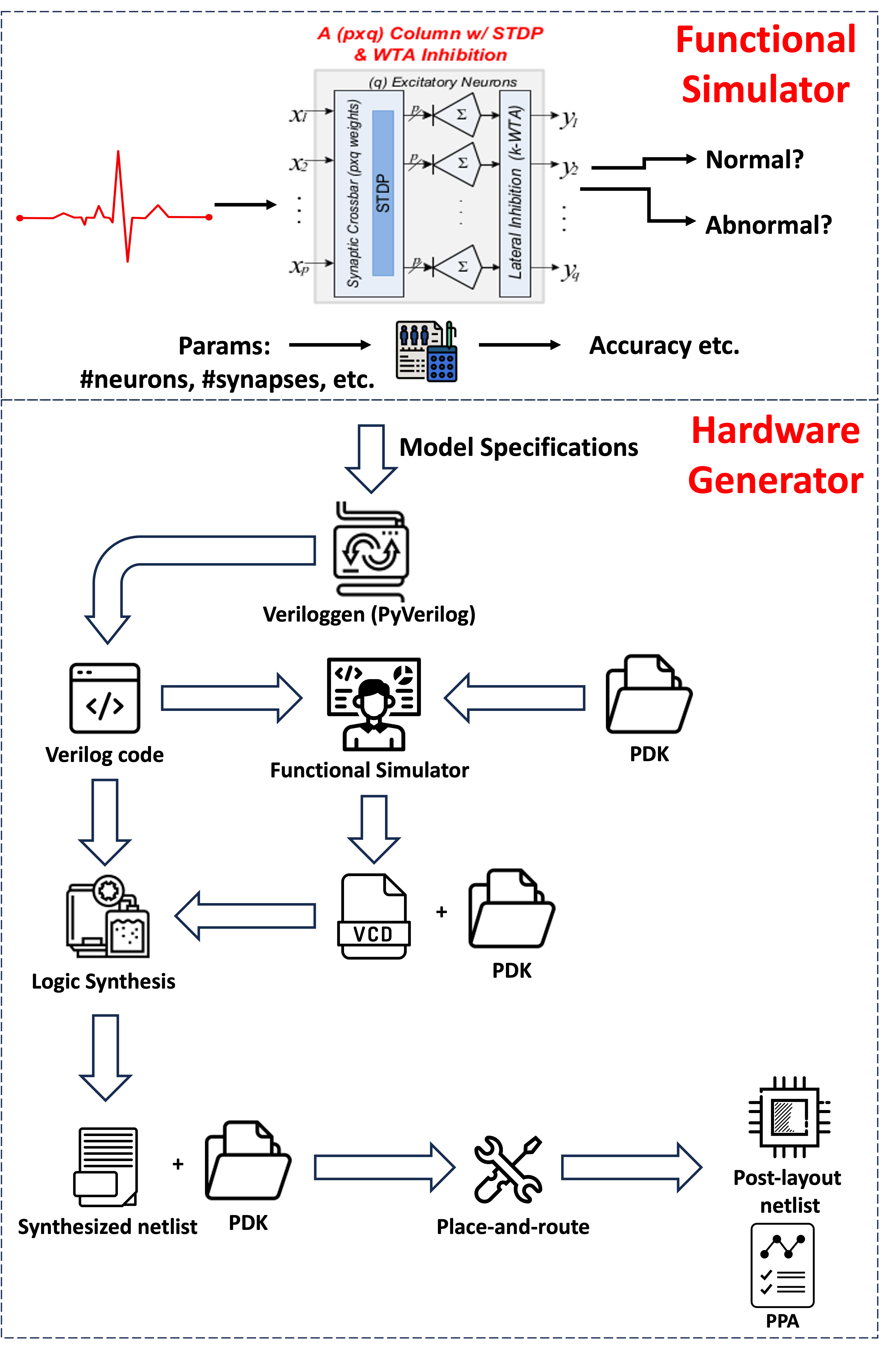} 
\caption{The TNNGen framework for designing TNN-based neuromorphic sensory processing units. It comprises a PyTorch functional simulator and a hardware generator, and automates the entire design flow  from PyTorch to chip layout.}
\label{tnngen_intro}
\end{figure}

This work serves as the first attempt at realizing such a framework for the automated design of application-specific TNNs, or \textit{Neuromorphic Sensory Processing Units} (NSPU), for online clustering of time-series sensory signals. As illustrated in Fig. \ref{tnngen_intro}, the framework leverages PyTorch \cite{paszke2019pytorch}, PyVerilog \cite{Takamaeda:2015:ARC:Pyverilog}, Python, Cadence toolchain, and open-source FreePDK45 \cite{oliveira2016ascend}, ASAP7, and TNN7 libraries \cite{nair2022tnn7}. Starting from high-level modeling of TNNs in PyTorch, this framework enables the generation of post-layout netlists of the models along with resulting hardware metrics in a single automated flow. It facilitates the development of optimized energy-efficient TNN designs without expert involvement, integrating previously segregated software-only and hardware-only TNN developments. Further, we enable users without EDA access to obtain key hardware results without running the  actual process flow, via \textit{forecasting}. Key contributions of this work are:

\begin{itemize}
    \item \textit{TNNGen} - a pioneering attempt at a design framework for automated design of custom TNN-based NSPUs.
    \item A novel TNNGen functional simulator based on PyTorch for robust modeling and rapid application exploration.
    \item A novel TNNGen hardware generator based on PyVerilog that translates TNN PyTorch models to layout in conjunction with Cadence EDA tools and a library of finely-tailored TCL scripts for optimizing the TNN designs.
    \item Time-series clustering performance and post-layout hardware metrics for seven representative TNN designs generated using TNNGen across different technology nodes, extending beyond previous post-synthesis studies.
    \item Accurate \textit{forecasting} of post-layout die area and leakage power for a quick evaluation of hardware complexity in lieu of the time-consuming hardware process flow.
\end{itemize}

The next section details the TNNGen framework and its key components. Section \ref{sec::result} describes our experimental setup and evaluation results, highlighting the optimized TNN designs generated and design flow runtimes, along with \textit{forecasting}.
Finally, Section \ref{sec::conclusion} summarizes key findings and future work. 

\section{TNNGen Design Framework}
\label{sec::tnngen}

The proposed TNNGen framework orchestrates the entire TNN design flow by
providing user-tunable parameters. It facilitates rapid TNN model development and application performance evaluation using a PyTorch-based simulator, which then provides the model specifications to an automated hardware flow that delivers highly optimized physical design netlists. 

\subsection{Functional Simulator}

A novel PyTorch-based functional simulator is developed as part of TNNGen for swift design space exploration and precise evaluation of application-specific metrics (e.g., classification accuracy, clustering rand index, F1 score, etc.). The simulator is flexible, offering users the ability to quickly explore a vast design space and use the resulting insights to develop optimized TNN models. Some key design space configurations include: (i) single-column TNNs with an arbitrary number of neurons ($q$) and synapses per neuron ($p$), and (ii) large multi-layer TNNs with an arbitrary number of layers and columns per layer with configurable inter-layer connectivity. It supports various neuron response function models (including step-no-leak, ramp-no-leak, leaky-integrate-and-fire), winner-take-all inhibition (with customizable winner count and tie-breaking options), and spike timing dependent plasticity (STDP) learning in both supervised and unsupervised modes. Pytorch's tensor operations are utilized to implement all the TNN functionalities for high simulation speed. Further, it also supports GPU acceleration through PyTorch's CUDA API.

TNNGen simulator models the time dimension precisely, aligning with the direct implementation methodology in \cite{nair2021microarchitecture} wherein spike times are dictated by precise hardware clock cycles. The simulator performs cycle-accurate temporal modeling for time windows around onset of spikes, and dynamically switches to an event-driven approach in time windows where spikes are absent to speed up the simulation. TNNGen employs a modular and parameterized approach, leveraging key functional blocks of TNNs. 

\subsection{Hardware Generator}

TNNGen automates the hardware design process flow by facilitating automated RTL generation, RTL simulation, logic synthesis, and place-and-route while ensuring a smooth  design flow. It leverages Veriloggen package built on top of PyVerilog \cite{Takamaeda:2015:ARC:Pyverilog} to provide a Python interface to the user for converting PyTorch model specifications 
to Verilog RTL codes.

Table \ref{tab:eda} specifies the process flows within TNNGen, the Cadence tools utilized during each flow, and the various libraries currently supported by the framework. Cadence EDA tools are specifically chosen as the ASAP7 and TNN7 libraries are primarily supported in the Cadence toolchain. However, TNNGen is built with huge focus on flexibility and modularity to enable easy integration of other toolchains and libraries. We plan to open-source TNNGen for the research community to not only leverage it for their custom TNN design flow but also enhance it with additional capabilities. 

In the TNNGen backend, to enable PyTorch-to-RTL conversion, we implemented all the TNN functionalities in PyVerilog, ensuring the generated RTL is highly optimized and aligns with the microarchitecture in \cite{nair2021microarchitecture}. TNN7 custom macros are incorporated to help accelerate runtime. \cite{nair2022tnn7} reports a 3x speedup for logic synthesis; we go a step further and report the place-and-route speedup in Section \ref{sec::result}. Further, TNNGen contains a library of specifically tailored TCL scripts and templates for automating the various design flows and PDKs, while providing end-user with complete freedom to configure the flow as needed.
Note that TNNGen does not cover DRC \& LVS checks for signoff as it requires expert intervention. 


\begin{table}[h]
    \centering
    \caption{Industry EDA tools supported by TNNGen.}
    \begin{tabular}{|c|c|}
        \hline
       \textbf{ Design Flow Stag}e & \textbf{Cadence Tool} \\
        \hline
        \hline
        RTL simulation  & Xcelium \\
        \hline
        Logic synthesis & Genus \\
        \hline
        Place-and-route & Innovus \\
        \hline
        \hline
        Library support & FreePDK45, ASAP7, TNN7 \\
        \hline
    \end{tabular}
    \label{tab:eda}
\end{table}

\begin{table}[t]
\centering
\caption{Seven different TNN configurations for various sensory modality applications used for the experimental setup. Clustering performance (rand index) for TNNs vs state-of-the-art DTCR, normalized to \textit{k}-means is also provided.}
\scalebox{0.9}{
 \begin{tabular}{|c|c|c|c|c|c|c|c|} 
 \hline
 UCR & UCR Benchmark & Sensory & DTCR & TNN \\
 Column & Name & Modality & Rand & Rand\\
 (\textit{p}x\textit{q}) &  &  & Index & Index\\
 \hline
 \hline
 65x2 & SonyAIBORobotSurface2 & Accelerometer & 0.8354 & 0.6066 \\
 \hline
 96x2 & ECG200 & ECG & 0.6648 & 0.6648\\
 \hline
 152x2 & Wafer & Fabrication process & 0.7338 & 0.555\\
 \hline
 343x2 & ToeSegmentation2 & Motion sensor & 0.8286 & 0.6683\\
 \hline
 637x2 & Lightning2 & Optical + RF sensor & 0.5913 & 0.577\\
 \hline
 470x5 & Beef & Food spectrograph & 0.8046 & 0.731\\
 \hline
 270x25 & WordSynonyms & 1D word outlines & 0.8984 & 0.8473\\
 \hline
 \end{tabular}
 }
  \label{Tab:column_config}
\end{table}


\begin{table}[t]
\centering
\caption{Post-place-and-route leakage power results for the seven representative UCR column designs.} 
\scalebox{0.94}{
 \begin{tabular}{|c|c|c|c|c|c|} 
 \hline
 UCR & Synapse & FreePDK45 & ASAP7 & TNN7 \\
 Benchmark  & Count & Leakage  & Leakage  & Leakage \\
 Name & & ($m$W) & ($\mu$W) & ($\mu$W) \\
 \hline
 \hline
 SonyAIBORobotSurface2 & 130 & 0.299 &  0.961  &  0.57 \\
 \hline
 ECG200 & 192 & 0.442  & 1.41 & 0.84 \\
 \hline
 Wafer & 304 & 0.717 & 2.26 &  1.34 \\
 \hline
 ToeSegmentation2 & 686 & 1.59 & 5.09 & 3.14 \\
 \hline
 Lightning2 & 1274 & 2.95 & 9.81 & 5.84\\
 \hline
Beef & 2350 & 5.452 &  17.4  & 11.06 \\
 \hline
 WordSynonyms & 6750 & 15.66 & 46.69 & 31.13 \\
 \hline
\end{tabular}
 }
  \label{tab:pnr_leakage}
\end{table}

\section{Results and Evaluation}
\label{sec::result}

\subsection{Experimental Setup}
We adopt the same seven single-column designs in \cite{shen2023cortical}, targeting different sensory modalities within the UCR archive \cite{UCRArchive2018}, as representative benchmarks to showcase TNN designs (Table \ref{Tab:column_config}). The designs are evaluated using three approaches:

\begin{itemize}
    \item Develop PyTorch TNN models as per $p$, $q$ parameters in Table \ref{Tab:column_config} and report corresponding clustering performance.
    \item Use TNNGen to generate post-layout hardware metrics across multiple cell libraries and technology nodes.
    \item Predict the hardware metrics without actually running any of the process flows, using TNNGen's \textit{forecasting} feature.
\end{itemize}

This research extends beyond previous post-synthesis studies on TNN implementations by presenting post-layout results using multiple cell libraries. Further, we provide layout runtime comparisons and assess the \textit{forecasting} feature of TNNGen that can predict post-layout hardware metrics without EDA runs. Our runtime simulations are run on a server with 8 Intel Xeon(R) E5-2680 CPU cores.

\subsection{TNNGen Design Performance and Hardware Complexity}

TNNGen simulator is used for modeling and rapidly simulating different single-column TNN designs targeting seven different sensory modalities. To evaluate the unsupervised clustering performance, \textit{rand index} is utilized, following the method outlined in \cite{chaudhari2021unsupervised}. Table \ref{Tab:column_config} shows the rand index results normalized to \textit{k}-means for TNN and a state-of-the-art deep learning algorithm DTCR \cite{ma2019learning}. The table shows that a single TNN column performs nearly as well as DTCR for four of the seven benchmarks but underperforms 
for the remaining three. On average, DTCR outperforms TNNs by nearly 12\%, aligning with the results in \cite{chaudhari2021unsupervised}.
It is essential to note that DTCR employs a significantly more complex DNN model, rendering it impractical for edge hardware deployment due to its computational demands. The simulator results demonstrate the efficacy of small TNN designs for time-series clustering.




\begin{table}[t]
\centering
\caption{Post-place-and-route  die area results for the seven representative UCR column designs.} 
\scalebox{0.93}{
 \begin{tabular}{|c|c|c|c|c|c|} 
 \hline
UCR & Synapse & FreePDK45 & ASAP7 & TNN7 \\
 Benchmark  & Count & Area & Area & Area \\
 Name & & ($\mu$m\textsuperscript{2}) & ($\mu$m\textsuperscript{2}) & ($\mu$m\textsuperscript{2}) \\
 \hline
 \hline
 SonyAIBORobotSurface2 & 130 & 14284.466 & 1028.67  & 692.06  \\
 \hline
 ECG200 & 192 & 21036.08 & 1513.05 & 1015.8 \\
 \hline
 Wafer & 304 & 33868.98  & 2394.01  & 1608.52 \\
 \hline
 ToeSegmentation2 & 686 & 75654.82 & 5388.72  &  3682.63 \\
 \hline
 Lightning2 & 1274 & 140,502.84 &  10184.45 & 6860.68 \\
 \hline
Beef & 2350 & 259,167.4 & 18298.1 & 12634.83  \\
 \hline
 WordSynonyms & 6750 & 744,422.4 & 51158.20  & 35303.88  \\
 \hline
 \end{tabular}
 }
  \label{tab:pnr_area}
\end{table}

\begin{figure}[t]
  \centering
  \subfloat[65x2 (79.2 ns)]{\includegraphics[width=0.158\textwidth]{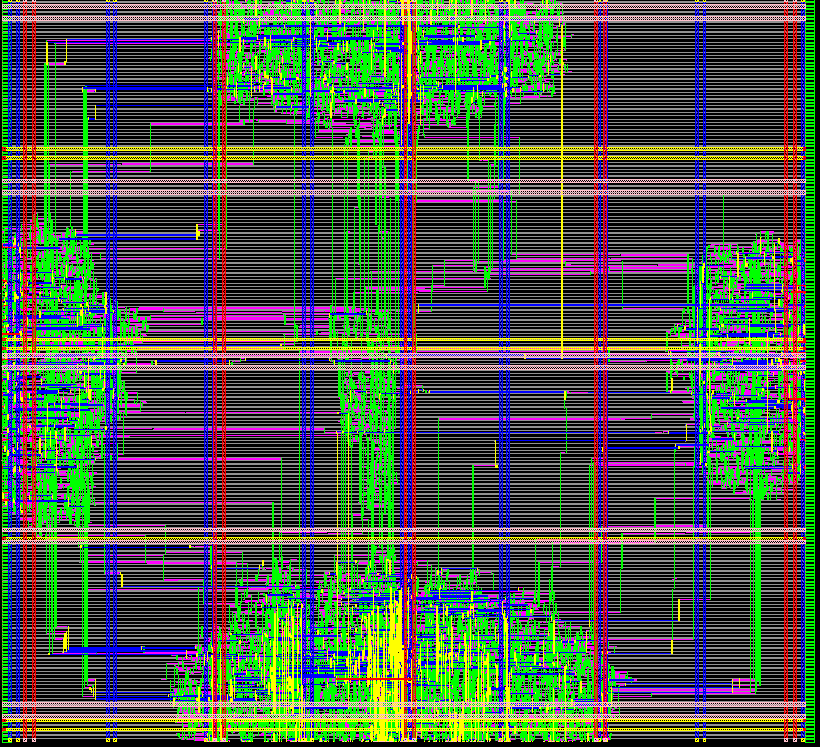}\label{fig:65x2}}
  \hfill
  \subfloat[96x2 (93.36 ns)]{\includegraphics[width=0.158\textwidth]{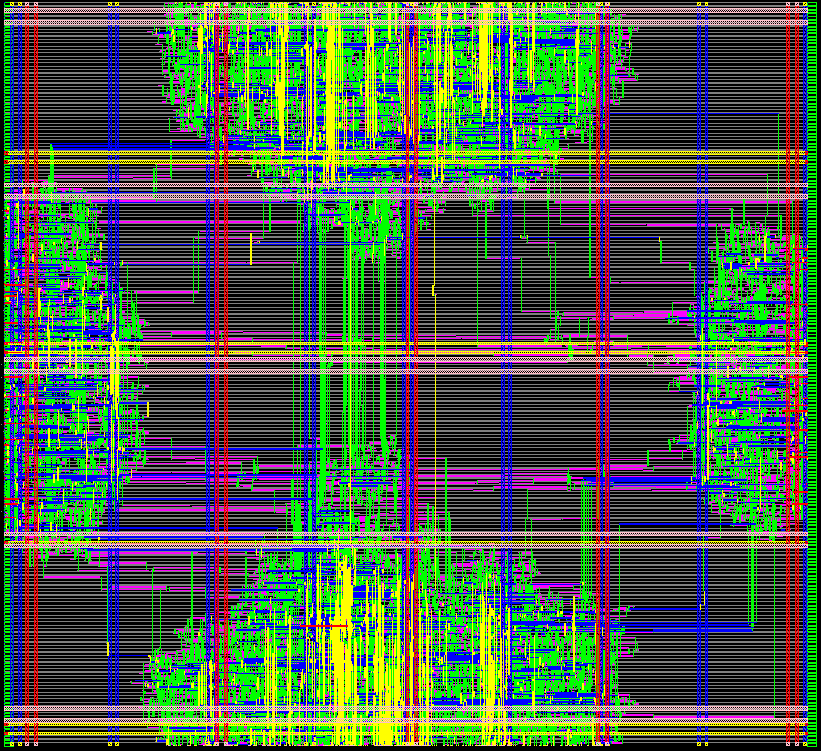}\label{fig:96x2}}
  \hfill
  \subfloat[152x2 (98.4 ns)]{\includegraphics[width=0.158\textwidth]{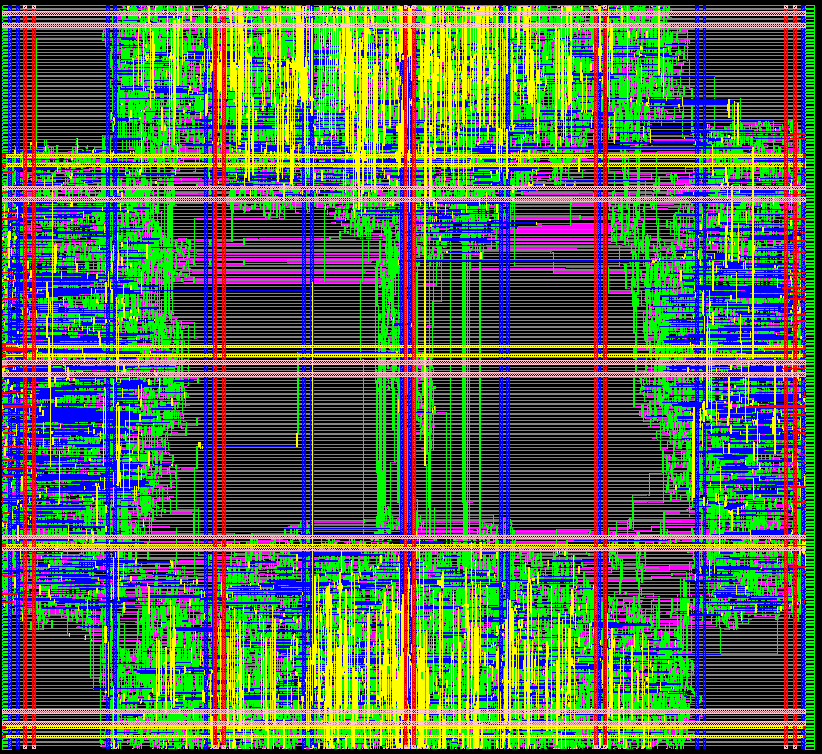}\label{fig:152x2}}
  \hfill
  \caption{Layouts of three generated column configurations fitted onto the same floorplan size. $p\times q$ column configurations are provided with computation latencies inside parentheses.}
  \label{fig:layouts}
\end{figure}

 TNNGen hardware generator translates the above software models to layouts.
 We employ various open cell libraries, with resulting hardware metrics reported in Tables \ref{tab:pnr_leakage} and \ref{tab:pnr_area}.
With TNN7, there is a 32.1\% and 38.6\% decrease in area and leakage power compared to ASAP7, respectively, aligning with the findings in \cite{nair2022tnn7}.
We report only leakage power here as total power requires fine-tuned physical rules specific to each design, including clock tree synthesis. Nevertheless, we do report the total power specifically for the largest column (6750 synapses) with TNN7 library for evaluation purposes. 

Using the TNN7 macros, the largest column results in just 0.035 mm\textsuperscript{2} area and consumes only 0.067 mW total (leakage + dynamic) power after layout, going beyond the post-synthesis area and total power reported in \cite{nair2022tnn7}. 
The corresponding FreePDK45 and ASAP7 area/leakage values are 0.744 mm\textsuperscript{2}/15.66 mW and 0.051 mm\textsuperscript{2}/0.047 mW respectively. 
The advantage of 7nm designs vs. 45nm is clear. We also see from Tables \ref{tab:pnr_leakage} and \ref{tab:pnr_area} that TNN7 (with custom macro cells) achieves better area and leakage than ASAP7.

For computation latency (i.e., per sample inference) evaluation, we first consider three smaller columns ($65 \times 2$, $96 \times 2$, $152 \times 2$) fitted for the same floorplan size, as shown in the layouts in Fig. \ref{fig:layouts}. The resulting computation times are 79.2 ns, 93.36 ns, and 98.4 ns, respectively. For the largest $270 \times 25$ column, the resulting latency is 180 ns. We can see that these TNN designs are extremely fast in performing inference and thereby ideal for low power real-time edge AI deployment.

\begin{figure}[t]
\centering
\includegraphics[width=0.95\columnwidth]{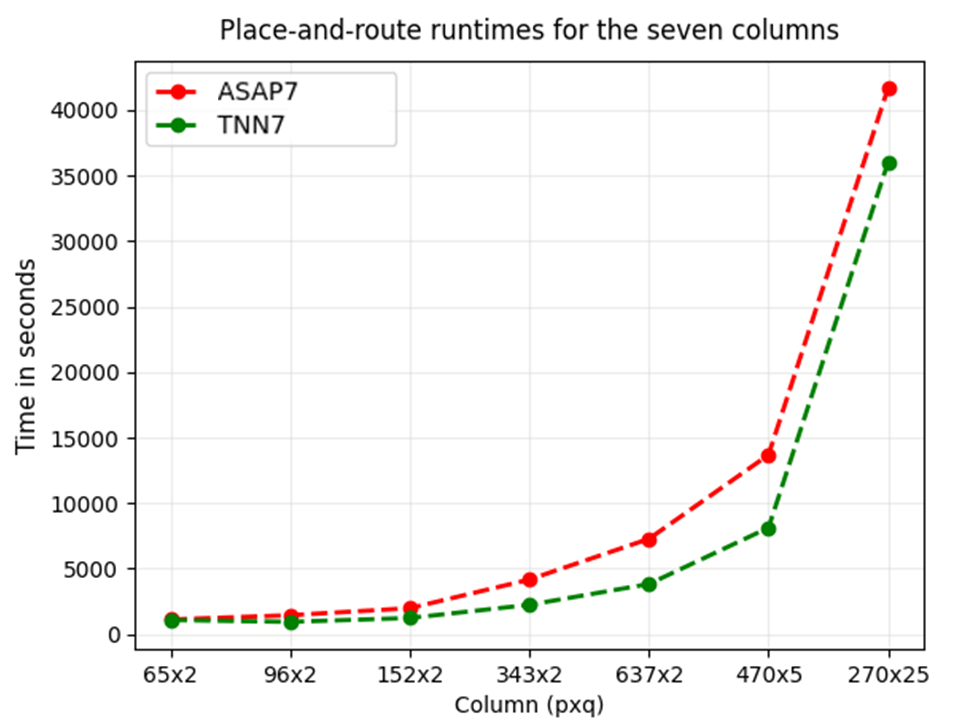} 
\caption{Innvous place-and-route runtime (in seconds) for baseline (ASAP7) and TNNGen (TNN7) column designs.}
\label{fig:runtime}
\end{figure}

\subsection{Runtime Evaluation}

Authors in \cite{nair2022tnn7} report an approximate 3x speedup during logic synthesis due to the use of TNN7 macros during the mapping and optimization phases. 
We further validate and extend their empirical results by taking a step further and evaluating runtime speedup during place-and-route using Innovus.

Fig. \ref{fig:runtime} illustrates the runtime for place-and-route for increasing column sizes 
using only ASAP7 cells vs. TNN7 macros. As depicted, runtime scales with increasing column sizes, but TNN7 macros yield a slower trend. On average, the layout runtime using TNN7 in Innovus place-and-route is roughly 32\% better than ASAP7. For the largest column, the entire hardware process flow (synthesis + place-and-route runtime) is reduced by almost 47\%, indicating larger designs benefit more in runtime speedup with TNN7's custom macros.

\begin{table}[t]
\centering
\caption{Forecasted (FC) post-place-and-route 7nm PPA for seven representative UCR column designs using TNNGen.}
\scalebox{0.94}{
 \begin{tabular}{|c|c|c|c|c|c|} 
 \hline
 UCR & Syn. & FC & FC & FC & FC \\
 Benchmark & Count & Area & Area & Leakage & Leakage\\
 Name &  & ($\mu m^{2}$) & Error & ($\mu W$) & Error\\
 \hline
 \hline
 SonyAIBORobot... & 130 & 627.9 & +10.36\% & - & - \\
 \hline
 ECG200 & 192 & 972.62 & +6.07\% & - & - \\
 \hline
 Wafer & 304 & 1595.34 & +2.25\% & 0.92 & +32.9\% \\
 \hline
 ToeSegmentation2 & 686 & 3719.26 & -0.33\% & 2.98 & +6.14\% \\
 \hline
 Lightning2 & 1274 & 6988.54 & -0.25\% & 6.16 & -1.72\% \\
 \hline
 Beef & 2350 & 12971.1 & -1.7\% & 11.98 & -5.1\% \\
 \hline
 WordSynonyms & 6750 & 37435.1 & +0.2\% & 35.77 & +0.52\% \\
 \hline
 \end{tabular}
 }
  \label{tab:fctable}
\end{table}
\subsection{Area and Leakage Power Forecasting}
Hardware development is typically time-consuming. Many researchers may not have access to commercial EDA tools. Hence, we integrate a \textit{forecasting} feature for predicting silicon die area based on TNN synapse count without actually running the TNNGen hardware flow.
This feature leverages the linear trends of area and leakage power with respect to total synapse count to build a linear regression model, which is trained on many TNNGen runs with varying TNN sizes.

\begin{figure}[t]
\centering
\includegraphics[width=0.9\columnwidth]{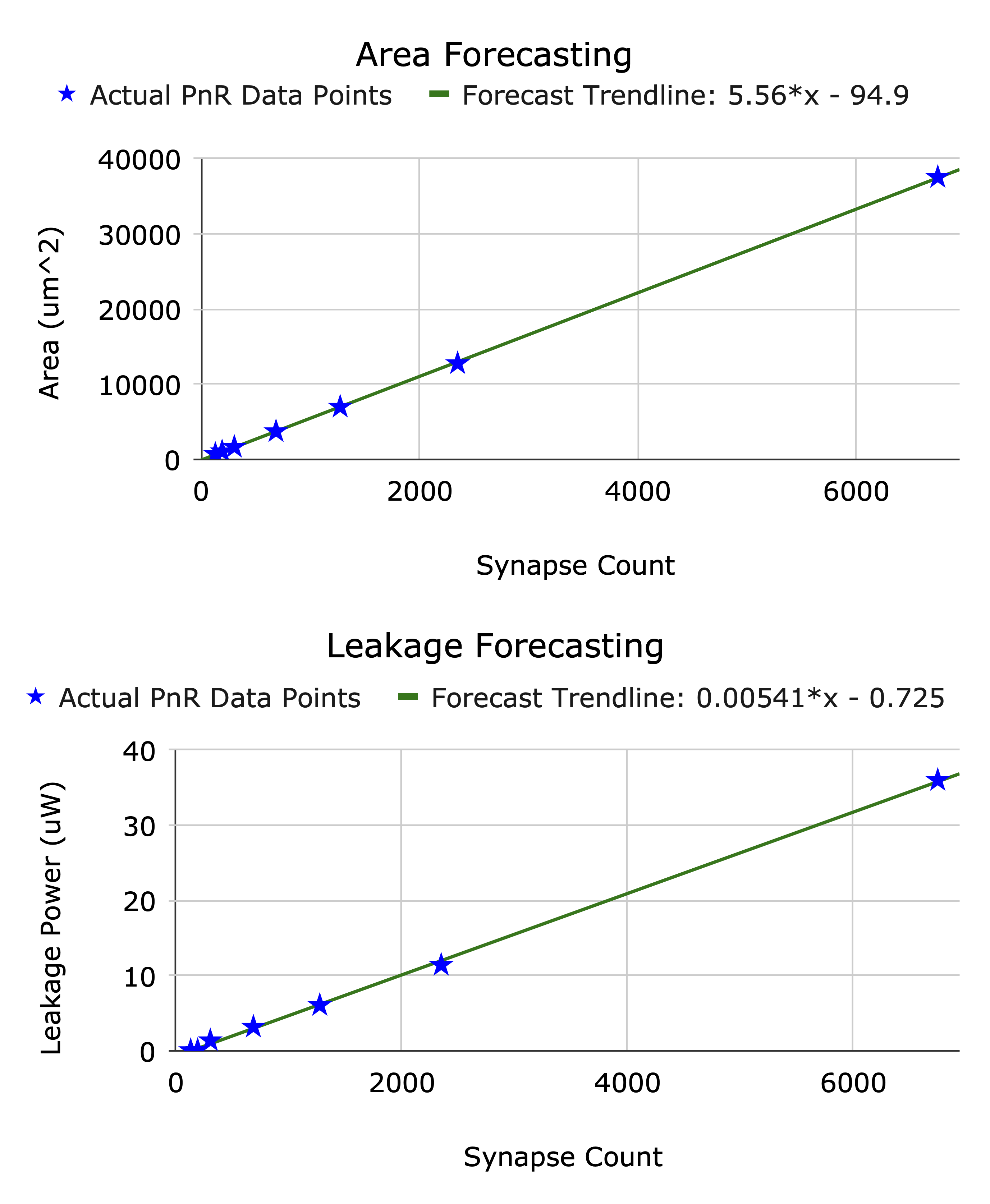} 
\caption{Area and Leakage power forecasting illustrating actual data points (`stars') and the forecasting trendline equations.}
\label{fig:leakagefc}
\end{figure}

The regression model for area follows the equation: \[Area=5.56*SynapseCount-94.9\] and leakage power follows the equation: \[Leakage=0.00541*SynapseCount-0.725\] Table \ref{tab:fctable} along with Fig. \ref{fig:leakagefc} report the forecasting (FC) results for area and leakage power, along with their forecasting errors. It can be seen that area can be predicted very accurately within 1\% of the original values for large designs. Leakage power, although inaccurate for small designs (omitted for the two smallest designs), is also highly accurate for large designs (the largest design only incurs 0.52\% error). Fig. \ref{fig:leakagefc} illustrates the efficacy of the linear trendline. The forecasting regression model is part of the TNNGen framework and can be continually refined with more actual design data points.



\section{Conclusion}
\label{sec::conclusion}
This paper serves as the first effort in creating an automated design framework (from PyTorch model to chip layout) for the design of application-specific TNN-based neuromorphic sensory processing units (NSPUs). TNNGen confirms the feasibility of such a framework.
Initial results indicate automated designs are highly efficient. Post-layout 7nm results show our largest benchmark design requires only 0.035 mm\textsuperscript{2} die area and 0.067 mW total power, with a compute latency of 180 ns.
This work also demonstrates the benefits of leveraging custom macros in improving both hardware metrics and design flow runtimes. We plan to develop a full library of custom macros that can be smoothly integrated into the framework. 
We also plan to extend the framework to support more diverse applications and much more complex multi-layer TNN designs.
The current framework stands as an important milestone for demonstrating the feasibility and effectiveness of an end-to-end toolchain for the automated design of application-specific TNNs for online  sensory signal processing. We plan to open source this framework to facilitate the experimentation and further enhancements by the research community.

\bibliographystyle{IEEEtranS}
\bibliography{refs}

\end{document}